\documentclass[conference]{IEEEtran}
\makeatletter
\def\ps@headings{%
\def\@oddhead{\mbox{}\scriptsize\rightmark \hfil \thepage}%
\def\@evenhead{\scriptsize\thepage \hfil \leftmark\mbox{}}%
\def\@oddfoot{}%
\def\@evenfoot{}}
\makeatother
\usepackage{amsmath}
\usepackage{amsfonts}
\usepackage{amssymb}
\usepackage{mathrsfs}
\usepackage{graphicx}
\usepackage{subfig}
\usepackage{setspace}

\newtheorem{theorem}{Theorem}[section]

\newtheorem{definition}[theorem]{Definition}
\newtheorem{problem}[theorem]{Problem}
\newtheorem{corollary}[theorem]{Corollary}

\newcommand{\N}{\mathbb{N}}

\newcommand{\BO}{\mathcal{O}}
\newcommand{\A}{\mathcal{A}}
\newcommand{\qed}{\hfill \ensuremath{\Box}}
\newcommand{\proof}[1]{\noindent\textnormal{\textbf{Proof. }}#1\qed}
\newcommand{\Ms}{\mathbf{S}}
\newcommand{\Mr}{\mathbf{R}}
\newcommand{\CX}{\mathrm{C}}

\title{CLEX: Yet Another Supercomputer Architecture?}

\author{\IEEEauthorblockN{Christoph Lenzen}
\IEEEauthorblockA{Department of Algorithms and Complexity\\
MPI for Informatics\\
Saarbr\"ucken, Germany\\
clenzen@mpi-inf.mpg.de}
\and
\IEEEauthorblockN{Roger Wattenhofer}
\IEEEauthorblockA{Computer Engineering and
Networks Laboratory (TIK)\\
ETH Zurich\\
Zurich, Switzerland\\
wattenhofer@ethz.ch}
}

\begin{document}

\maketitle

\begin{abstract}
We propose the CLEX supercomputer topology and routing scheme. We prove that CLEX
can utilize a constant fraction of the total bandwidth for point-to-point
communication, at delays proportional to the sum of the number of intermediate
hops and the maximum physical distance between any two nodes. Moreover,
all-to-all communication can be realized $(1+o(1))$-optimally both with regard to
bandwidth and delays. This is achieved at node degrees of $n^{\varepsilon}$, for
an arbitrary small constant $\varepsilon\in (0,1]$. In contrast, these results
are impossible in any network featuring constant or polylogarithmic node degrees.
Through simulation, we assess the benefits of an implementation of the proposed
communication strategy. Our results indicate that, for a million processors, CLEX
can increase bandwidth utilization and reduce average routing path length by at
least factors $10$ respectively $5$ in comparison to a torus network.
Furthermore, the CLEX communication scheme features several other properties,
such as deadlock-freedom, inherent fault-tolerance, and canonical partition into
smaller subsystems.
\end{abstract}




\section{Introduction \& Related Work}

Ever since the advent of massively parallel computing architectures, there has
been lively interest in the question how the nodes\footnote{By ``node'' we mean
the smallest computing unit that can be seen as (more or less) a sequentially
working device. In today's multiprocessor systems this means a single core.} of
a supercomputer should be interconnected, e.g.\ as a fat tree, butterfly, or
hypercube~\cite{adams87,ahn09,goodman81,greenberg89,guo08,johnson89,kim07,
leighton92,leiserson96,leonard07,raghunath93,siegel89,zhu08}. Naturally, these
topologies try to balance between the desires for small node degrees and short
routing paths. Moreover, it is crucial to serve routing~requests~in~parallel,
using little, distributed computation, and to deal with faults. Eventually,
the theoretical understanding of these issues became systematic and
mature~\cite{heydemann97,shamir89,upfal92,even97}.

Today, communication in supercomputers is by and large implemented by means of
low-degree interconnection topologies. Thus, one would expect to find the
well-analysed topologies that have been proposed decades ago to dominate the
market. But far from it! State-of-the-art architectures like Cray XMT or IBM
Blue Gene provide point-to-point communication on top of a three-dimensional
torus network~\cite{adiga05,konecny07}. This appealing simplicity in design
comes at a cost, as such a system is fundamentally limited in communication.
Konecny~\cite{konecny07} writes ``Because the most interesting mode of operation
assumes uniformly distributed traffic, the network performance is expected to be
dominated by the bisection bandwidth.'' In a three-dimensional torus of $n=k^3$
processors, one can partition the processors such that two subsets of $n/2$
processors are connected by $2k^2$ edges only. In other words, because of
communication limitations ``the third dimension'' of processing is lost, since
the average point-to-point bandwidth between these subsets scales with $2/k$
times the individual link capacity. Bluntly, in today's supercomputers, for $n
\approx 10^6$ processors the torus architecture restricts communication to less
than 1\% of the total available bandwidth in the worst case.


So why is it that such an apparently suboptimal design is chosen by
practitioners? We believe the answer to this question to be twofold. On the one
hand, a (locally) grid-like communication network is of course well-suited to
deal with communication patterns that are local as well.\footnote{This is for
instance true for computational problems arising from physical systems, e.g.,
from fluid or solid body dynamics, which have been a (if not the) main focus of
parallel computing in the past.} We argue, however, that
this approach has several shortcomings. Firstly, it restricts the range of
problems for which the computer architecture is fitting to problems that are
parallelizable in a way that matches the network topology. Secondly, programmers
need to be aware of this issue and program accordingly, which might be a
non-trivial and error-prone task. Thirdly, on large scales, load balancing
issues may result in more complex, less local communication patterns if an
efficient progress of computation is to be ensured. And finally, even if all
these issues can be overcome at the time when the system goes online, it will
typically be in use for several years, implying that it is difficult to predict
whether demands will change during the life-time of the supercomputer.

On the other hand, the theory on interconnection networks fails to address some
questions of practical significance. For one, how should one actually
\emph{realize} one of the suggested topologies? This turns out to be critical
for performance, as the efficiency of the whole communication infrastructure
might break down because some of the physical links are exceedingly long: these
connections will suffer larger communication delays, consume more space and
energy, and complicate the physical layout of the system. To the best of our
knowledge, this issue has been neglected in all theoretical studies of the
matter; in stark contrast, even for the three-dimensional torus, which is fairly
amenable to low-distortion ``embedding'', optimizing the stretch has been
considered a worthwhile task~\cite{yu06}. What is more, we believe it to be
important to devise routing algorithms that deal with faults in a seamless,
fully distributed, and automatic manner. Therefore, it is not enough to show
that a topology exhibits a large number of disjoint short path between to
destinations, but one also needs to give a routing scheme that exploits the high
connectivity of the system to establish robustness with respect to failing nodes
or links.


In consequence, we would like to revive the interconnection discussion from a
theoretical point of view\footnote{Numerous works are published all the time,
but typically a topology is chosen and tested using standard routing mechanisms.
For instance, \cite{kim08} provides a two-level architecture similar to a
two-level CLEX system, but no mated routing algorithms or theoretical analysis
is given.} by presenting a new topology we call CLEX (CLique-EXpander).
Essentially, the CLEX design is the result of seeking
efficient communication in a world of physical constraints. To this end, we
deviate from standard analysis by measuring delays not solely in terms of hops,
but also considering the physical distance a signal needs to
travel.\footnote{Although typically bandwidth is the primary concern, recently
Barroso pointed out that it is feasible and crucial to strive for small delays
in warehouse-scale computing~\cite{barroso11}.} We prove that a point-to-point
communication bandwidth per node (to and from arbitrary destinations) matching
the total bandwidth per node up to a constant factor can be achieved, at delays
that are (asymptotically) proportional to the maximum physical distance between
any two nodes. Moreover, applying an asymmetric bandwidth assignment to the
links, all-to-all communication\footnote{Adiga et al.~\cite{adiga05} state that
``MPI\_AlltoAll is an important MPI collective communications operation in which
every node sends a different message to every other node.''} can be realized
$(1+o(1))$-optimally both with regard to bandwidth and delays.

As constant or polylogarithmic node degrees necessarily incur an average hop
distance of $\Omega(\log n/\log \log n)$, the price we pay for these properties
are node degrees of $n^{\varepsilon}$, for an arbitrarily small constant
$\varepsilon$. However, these fairly high degrees are ``localized'' in the sense
that all but a constant number of them connect the nodes of the basic building
blocks of our topology, i.e., cliques of size $n^{\varepsilon}$. Thus, one way
to interpret our results is to view the CLEX approach as a method to localize
the issue of an efficient (low-degree) communication network to much smaller
systems of $n^{\varepsilon}$ nodes, which may e.g.\ reside on a single
multi-core board. A multi-core board will offer means of on-board communication
by itself, and due to small distances and integrated circuits one can expect it
to be of greater efficiency than that of a comparable large-scale network. Thus,
the high connectivity of a CLEX system could be considered an abstraction that
can be replaced by any efficient local communication scheme within the cliques
(cf.~e.g.~\cite{kim07}).

Nonetheless, we do also propose a routing scheme that indeed is designed for the
high-degree CLEX network as is. Within cliques, it employs recent results on
parallel randomized load balancing~\cite{lenzen11}, ensuring a high degree of
efficiency and resilience of the overall approach. From our point of view, the
properties of the resulting system justify to re-raise the question whether high
degrees can be worth the effort. In fact, one could see this as another step of
localization: Our algorithm reduces the routing problem on the clique level to
one on the node level, namely to the one of efficiently routing between
$n^{\varepsilon}$ input and output ports. This task now is to be solved on a
physically much smaller scale, dealing with smaller communication delays and
being able to rely on much better synchronization between the individual
components. Again, one is free to replace the full connectivity between the
ports by any combination of topology and routing scheme that is efficient at
this scale.

To add some salt to the above theoretical considerations, we assess the
efficiency of a CLEX architecture in practice. To this end, we simulate
point-to-point communication in two systems comprising $32^4\approx 1,000,000$
nodes and $64^3\approx 250,000$ nodes.
The results of our simulation indicate that the usable bandwidth of a CLEX
architecture could be an order of magnitude larger than the \emph{theoretical}
optimum of the IBM Blue Gene and Cray XMT tori interconnection networks. Since
our comparison assumes identical total bandwidth in both designs, this is not a mere
consequence of indirectly increasing bandwidth via node degrees, but a
fundamental difference of the underlying topologies.


\section{Topology and Routing Algorithms}
In this section, we give solutions to the all-to-all and point-to-point
communication problems. To this end, we define an abstract model amenable to
formal analysis. However, the applied proof techniques extend to stronger models
which better match a real-world system. In particular, the assumptions of
asynchronicity and fault-free behaviour can be dropped. After describing the
topology of the CLEX architecture, we briefly compare two algorithms solving
all-to-all communication efficiently on our topology and the
three-dimensional torus. Finally, we give an algorithm for point-to-point
communication and analyze its synchronous running time. Our theoretical
findings are supported by the simulations presented in
Section~\ref{sec:simulation}.

\subsection{Model and Problem Formulation}
We model a supercomputer as an undirected graph $G=(V,E)$, $n:=|V|$, where nodes
represent the computing elements (processors) and edges bidirectional
communication links. To simplify the presentation, we assume that for each
$v\in V$, the loop $\{v,v\}$ is contained in $E$, i.e., nodes may ``send
messages to themselves''. We assume that communication is reliable and proceeds
in synchronous rounds.
Message size is in $\Omega(\log n)$, i.e., we assume that a constant number of
node identifiers of size $\log n$ fits into a message. Any upper bound on the
message size respecting this constraint is feasible; for the purpose of our
analysis, we however assume that in each round only one ``unit payload'' can be
sent by each node along each edge. Nodes have access to an infinite source of
random bits.
We point out, however, that our algorithms
will in practice work reliably also with pseudo-random instead of true random bits,
since all our results hold with high probability (w.h.p.)\footnote{That is,
with probability at least $1-1/n^c$ for a tunable constant $c>0$.}

Note that the assumptions on the communication model (which simplify the
presentation) can be considerably relaxed. Our algorithms can be run
asynchronously by including round counters into messages. Furthermore, as
demonstrated in~\cite{lenzen11}, the load balancing scheme can be made resilient
to a constant (independent) probability of message loss.

Observe that the total delay a message suffers comprises the time it takes being
relayed at intermediate nodes plus the time the signal travels along the edges
of the interconnection network. Thus, the simplistic measure given by round
complexity may not be accurate in practice; we also need to understand the
influence of propagation times. Therefore, we define the maximal (average) delay
$d$ ($\bar{d}$) as
\begin{equation*}
d := c_h h + c_p p\quad
\mbox{and}\quad\bar{d} := c_h \bar{h} + c_p \bar{p},
\end{equation*}
where $h$ ($\bar{h}$) is the maximal (average) number of hops until a message is
delivered, $p$ ($\bar{p}$) is the maximal (average) physical distance a message
travels, and $c_h$ and $c_p$ are appropriate constants (comprising units). To
get an idea of the order of magnitude of the respective terms, suppose that it
takes a few clock cycles before a message can be relayed (to a free channel) and
clock speeds are in the order of gigahertz. Thus, forwarding a message is
initiated after a few nanoseconds. At speed of light, a signal travels about a
foot per nanosecond.

Formally, we will solve the following problems.
\begin{problem}[Point-to-Point Communication]\label{prob:idt}Each node $v$ is
given a (finite) set of messages
\begin{equation*}
{\cal S}_v=\{m_v^i\,|\,i\in \{1,\ldots,i_v^{\max}\}\}
\end{equation*}
with destinations $d(m_v^i)\in V$. The goal is to deliver all messages to their
destinations, minimizing delays. By
\begin{equation*}
{\cal R}_v:=\big\{m_w^i\in \cup_{w\in V}{\cal
S}_w\,\big|\,d(m_w^i)=v\big\}
\end{equation*}
we denote the set of messages a node $v\in V$ shall receive. We abbreviate
$\Ms:=\max_{v\in V}|{\cal S}_v|$ and $\Mr:=\max_{v\in V}|{\cal R}_v|$, i.e.,
the maximal numbers of messages a single node needs to send or receive,
respectively.
\end{problem}

All-to-all communication is a special case of point-to-point communication.
\begin{problem}[All-To-All Communication]\label{prob:ata}
Each node $v\in V$ is given a message $m_v$. The goal is to deliver (a copy
of) each message $m_v$ to all nodes, minimizing delay.
\end{problem}

Note that this problem is easier to solve, since by setting ${\cal
S}_v=\{m_v^w\,|\,w\in V\}$ and $d(m_v^w)=w$ an instance of
Problem~\ref{prob:idt} is obtained.

\subsection{Interconnection Network}\label{sec:network}
Evidently, with node degrees of at most $\Delta$, any algorithm for
Problem~\ref{prob:ata} must take at least $n/\Delta$ rounds to complete.
Similarly, Problem~\ref{prob:idt} cannot be solved in less than
$\max\{\lceil{\Ms}/\Delta\rceil,\lceil{\Mr}/\Delta\rceil\}$ rounds, as no node
can send or receive more than $\Delta$ messages in each round. Thus, in order to
hope for good running times, the communication graph needs to expand very
quickly, i.e., for any set of nodes $S\subset V$ with $|S|\leq n/2$ it is
necessary that $S$ has $\Omega(\Delta |S|)$ outgoing edges. At the same time, we
need to be aware that long-range links bridging a large physical distance should
not be used frequently, which needs to be respected by our routing scheme and
thus also the underlying topology. This motivates the following recursive graph
construction.

\begin{figure}[tb!]
\begin{center}
  \begin{subfloat}
  \centering
  \includegraphics[width=.7\columnwidth]{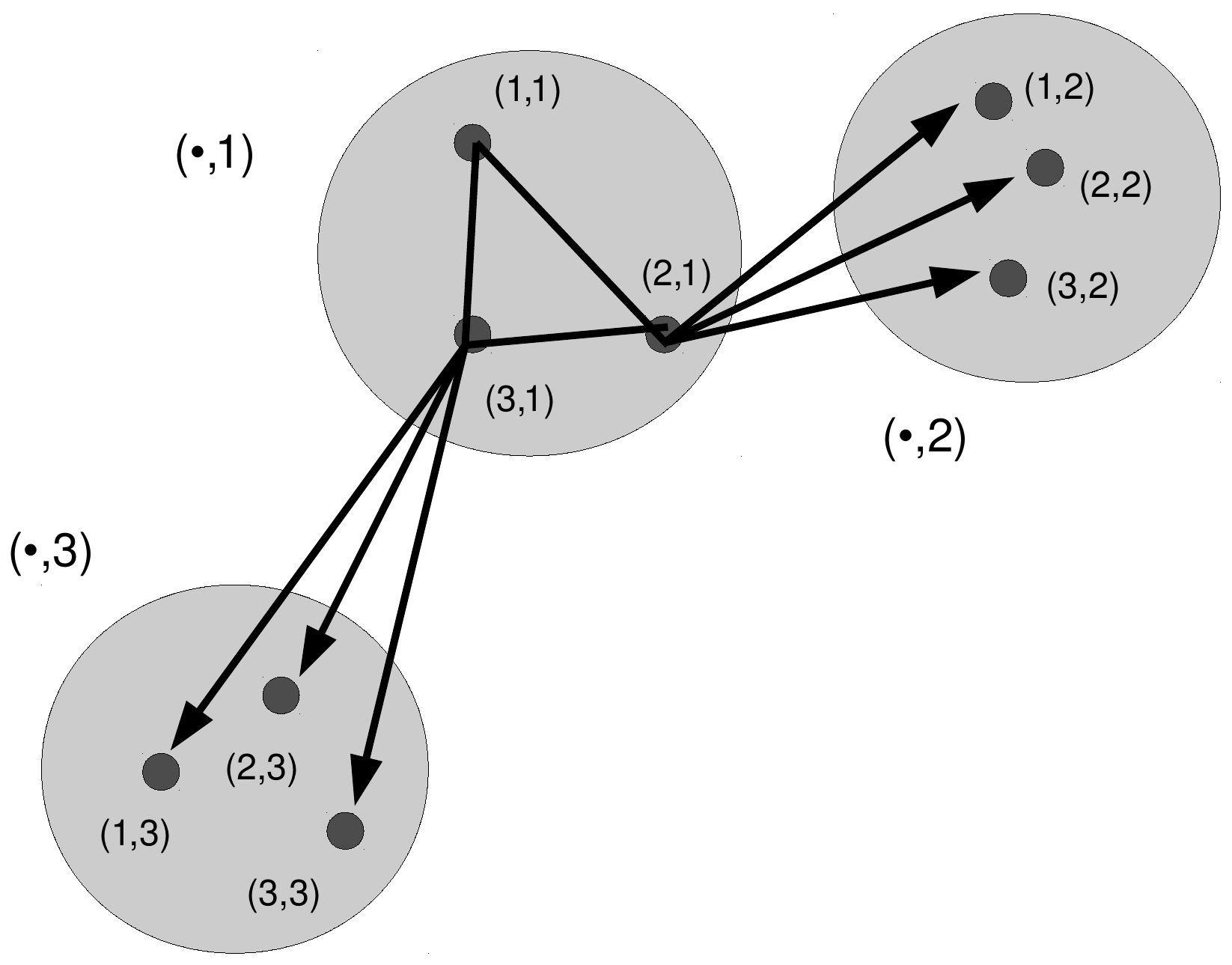}
  \caption{Illustration of $C(1/3,2)$. Each copy of $C(1/3,1)$ is
  enclosed in a grey circle. The links without arrowheads are bidirectional. For
  clarity, only the outgoing links of the nodes in the center copy are depicted,
  and the redundant links of node $(1,1)$ on level $1$ (connecting to the nodes
  of the first copy) are left out.}\label{fig:clex_2}
  \end{subfloat}
  
  \begin{subfloat}
  \centering
  \includegraphics[width=.7\columnwidth]{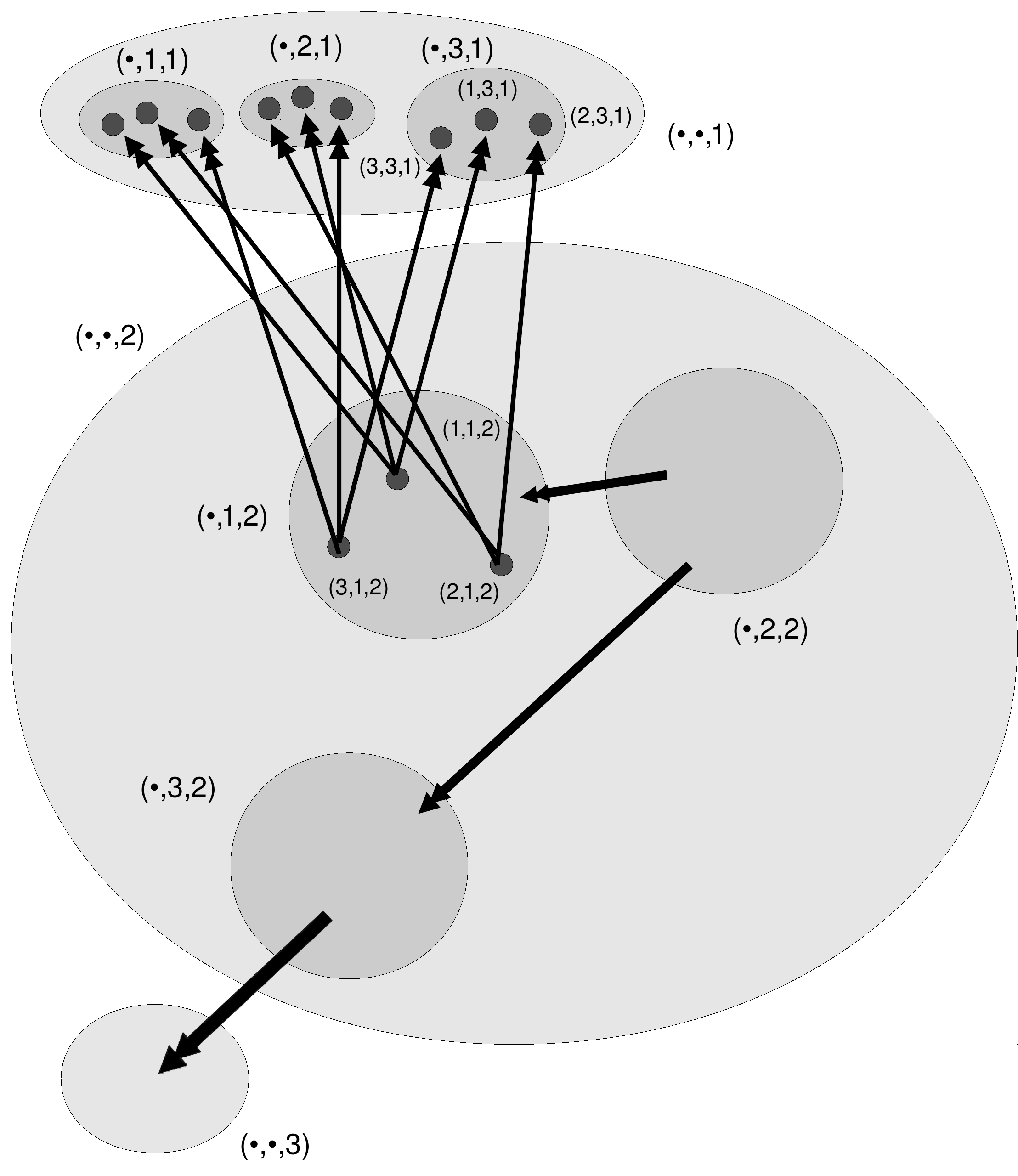}
  \caption{Illustration of the level $2$ links of $C(1/3,3)$. Copies of
  $C(1/3,2)$ and $C(1/3,1)$ are enclosed by light and dark grey circles,
  respectively. Only the outgoing links of nodes in the center copy of
  $C(1/3,1)$ are depicted. The remaining arrows subsume the connections of the
  copies of $C(1/3,1)$ labeled $(\cdot,2,2)$ and $(\cdot,3,2)$. Note that it
  is not necessary to connect the subgraphs on level 2 or higher precisely in
  this way, as long as bandwidths are well-balanced on each level.}
  \label{fig:clex}
  \end{subfloat}
\end{center}
\vspace*{-.7cm}
\end{figure}

\begin{definition}[CLEX Graphs]
Suppose for a constant $s\in (0,1]$ that $n^s$ and $1/s$ are integer. We
recursively define the (directed) \emph{CLEX graph} $\CX(s,l)$ of $l\in
\{1,\ldots,1/s\}$ \emph{levels}. Set $\CX(s,1):=K_{n^s}$, i.e., a clique of
$n^s$ nodes, and label its nodes $(1),(2),\ldots,(n^s)$. Assuming that
$\CX(s,l)$ is already defined, $\CX(s,l+1)$ is composed of $n^s$ isomorphic
copies $\CX(s,l)_i$, $i\in \{1,\ldots,n^s\}$, plus additional edges. Using the
label $(v_1,\ldots,v_l)\in \{1,\ldots,n^s\}^l$ a node $v\in V(\CX(s,l))_i$
inherits from $\CX(s,l)_i$, we can identify it uniquely with
$(v_1,\ldots,v_l,i)\in\{1,\ldots,n^s\}^{l+1}$.

The edges of $\CX(s,l+1)$ are all edges contained in the $\CX(s,l)_i$ plus
\begin{eqnarray*}
E_{i,l+1}&:=&\{((v_1,\ldots,v_l,i),(v_1,\ldots,v_{l-1},j,v_l))\\
&&~|\,j\in \{1,\ldots,n^s\} \wedge v\in V(\CX(s,l)_i)\},
\end{eqnarray*}
i.e., $E(\CX(s,l+1))=\cup_{i=1}^{n^s} (E(\CX(s,l)_i)\cup E_{i,l+1})$. See
Figures~\ref{fig:clex_2} and~\ref{fig:clex} for an illustration.
\end{definition}

Observe that each copy of $\CX(s,l+1)$ connects each of its subgraphs
$\CX(s,l)_i$ by $|V(\CX(s,l))|$ many edges to any $\CX(s,l)_j$, $j\in
1,\ldots,n^s$, such that degrees increase by exactly $n^s$ on each level. Thus,
$\CX(s,1/s)$ has uniform degrees of $n^s/s-1$. Its diameter $D(\CX(s,1/s))$ is
bounded by $2^s-1$, as $D(K_{n_s})=1$ and $D(\CX(s,l+1))$ is at most
$2D(\CX(s,l))+1$.\footnote{Any copy of $\CX(s,l)$ is connected to all other
copies, hence we can follow a shortest path in $\CX(s,l)$ to one endpoint of
this edge, traverse it, and follow another shortest path in $\CX(s,l)$ to the
destination.} We remark that these graphs are not Cayley graphs
(cf.~\cite{heydemann97}).

Note that if for any $k,t\in \N$ we set $n:=k^t$ and $s:=1/t$, both $1/s$ and
$n^s$ are integer, i.e., $\CX(s,l)$ is defined for $l\in \{1,\ldots,t\}$. In
other words, it is possible to choose $s$ arbitrarily small and $n$ arbitrarily
large.

The CLEX topology can be realized with e.g.\ a grid-like node positioning. We
ensure that nodes that are connected on low levels are close to each other by
arranging them in cubes. On each new level, we simply arrange an appropriate
number of cubes to a larger cube. Such a system will not experience a
significant stretch in distances due to embedding issues.

As will emerge from the analysis, it is feasible to replace nodes' $n^s$ edges
in $E_{i,l+1}$ by a single link of capacity $n^s$ to one of the endpoints of
these edges (such that each node gets also exactly one incoming link of this
capacity). This is also compatible with the solution of Problem~\ref{prob:ata}
proposed in Section~\ref{subsec:all-to-all}. Note that this way, node degrees
become $n^s+1/s-2$, with merely $1/s-1$ long-range links (i.e., links that are
not on the basic level). Clearly, this is to be preferred in any real-world
system, however, for ease of presentation, we stick to uniform edge capacities
in our exposition.

\subsection{All-to-All Communication}\label{subsec:all-to-all}
Problem~\ref{prob:ata} has simple solutions both on the torus and on CLEX
graphs. On the torus, first nodes exchange all messages in $x$-direction, then
$y$-, and finally $z$-direction. This defines for each message a tree with the
source as root on which the message is flooded. Thus this scheme is
bandwidth-optimal up to factor three. If links are congested, i.e., the
necessary traffic exceeds the available bandwidth, asymptotic optimality with
regard to delays follows from this observation. On the other hand, in absence of
congestion the solution is also delay-optimal, as the trees have minimal depth.
For the CLEX design, things are less obvious.

We will show that physical average and maximal routing path lengths can also be
kept close to the optimum in CLEX systems. For the sake of simplicity,
throughout this paper we assume that the torus is (locally) a perfect
three-dimensional grid of $k_1\cdot k_2\cdot k_3$ nodes.\footnote{Due to
physical constraints, the embedding will incur an additional stretch
(cf.~\cite{yu06}).} We assume that the CLEX topology is realized in a
hierarchical cube structure as described in Section~\ref{sec:network}.
We assume that cable connections are as short as possible, for both considered
topologies.\footnote{However, one might want to arrange connections in a CLEX
system in a more convenient manner, resulting in a small increase in cable
length whose influence we neglect.}

Using the previously explained scheme to solve Problem~\ref{prob:ata}
on the torus, messages travel on average $\bar{h}^T=(k_1+k_2+k_3)/2\geq
3n^{1/3}/2$ many hops. Hence, the maximal delay in an uncongested setting would
be roughly proportional to this value. Observe that no architecture can perform
significantly better, as processors cannot be packed much more densely because
of cooling issues and physical routing path lengths are optimal up to a factor
of $\sqrt{3}$.

The strategy to solve Problem~\ref{prob:ata} on $\CX(s,1/s)$ is very similar to
the one for tori. Each message $m_v$ is flooded along a tree induced by with
respect to hop distance shortest paths from $v$ to all nodes, where links on
lower levels are preferred (because they bridge shorter distances), giving a
bandwidth-optimal solution up to factor $1/s$, since links on level one have to
deal with most of the load. Note that an asymmetric bandwidth assignment to the
different levels reduces this factor, cf.~Section~\ref{subsec:dense}. Messages
are delivered to all destinations after travelling at most one edge on each
level. Since we assumed that processors are arranged in a cubic grid and links
are direct connections, maximal link lengths on Level~$l\in \{1,\ldots,1/s\}$
are $\sqrt{3}n^{ls/3}/2$, i.e., maximal propagation delays are
\begin{equation*}
c_p\bar{p}=c_p\frac{\sqrt{3}n^{1/3}}{2}\sum_{i=0}^{1/s-1}n^{-is/3}
\in c_p\frac{(1+o(1))\sqrt{3}n^{1/3}}{2}.
\end{equation*}
Hence, we achieve a $(1+o(1))$-approximation to physically optimal delays in
$C(s,1/s)$, which on three-dimensional tori is impossible. For the test
settings presented in Section~\ref{sec:simulation}, the $(1+o(1))$ term is
close to $\sqrt{3}$, i.e., the algorithm will at least perform as good as any
solution on a torus interconnection network with regard to propagation delays.

Moreover, as observed in Section~\ref{sec:network}, the diameter of
$\CX(s,1/s)$ is $2^{1/s}-1$, i.e., messages make at most that many hops. For
fixed $s$, we thus achieve asymptotically optimal maximal delays $d$
proportional to the maximal spatial distance between any two nodes. For the
parameter values considered in Section~\ref{sec:simulation}, the number of hops
reduces about a factor $10$ in comparison to a torus network of the same size.
As in a torus network typically the number of hops will be the dominant factor
contributing to delays in all-to-all communication, a CLEX network promises a
considerable improvement.

\subsection{Point-to-Point Communication}\label{sec:point-to-point}
In the following, w.l.o.g.\ we assume that in Problem~\ref{prob:idt} $\Ms=\Mr$,
as e.g.\ in case $\Ms<\Mr$ the number of messages each node needs to send is
upper bounded by $\Mr$ (and we want to show a bound that is essentially linear
in $\Ms+\Mr$). Rerouting each message through a uniformly and independently at
random (u.i.r.) chosen intermediate node (i.e., applying Valiant's
trick~\cite{valiant82}), we need to solve the two slightly simpler problems that
$(i)$ each node needs to send at most $\Ms$ messages whose destinations are
distributed u.i.r.\ or $(ii)$ each node needs to receive at most $\Mr$ messages
whose origins are distributed u.i.r. Note that these problems are
(asymptotically speaking) indeed less difficult, as applying Chernoff's bound we
see that w.h.p., in the first case each node needs to receive at most $\Mr'\in
(1+o(1))\Ms$ messages, while in the second case, no node initially holds more
than $\Ms'=\Mr'$ many messages. Thus, for the sake of analyzing the asymptotic
complexity of the problem, w.l.o.g.\ we assume in the following that message
destinations are distributed u.i.r.

We proceed by defining and analyzing the Algorithms $\A(l)$, $l\in
\{1,\ldots,1/s\}$, which solve Problem~\ref{prob:idt} on $\CX(s,l)$.
In case of $l=1$, the communication graph is simply $K_{n^s}$, thus we can use
an algorithm suitable for complete graphs. The advantage of full connectivity is
that any node may serve as relay for any message, reducing the routing problem
to a load balancing task. We follow the approach from~\cite{lenzen11}. For the
sake of clarity, we present a simplified algorithm to illustrate the concept.
Initialize $i:=1$ and $k(1):=1$ at each node. The algorithm executes the
following loop until all messages are delivered:
\begin{enumerate}
  \item Create $\lfloor k(i)\rfloor$ copies of each message. Distribute these
  copies uniformly at random among all nodes, but under the constraint that (up
  to one) all nodes receive the same number of messages.
  \item To each node, forward one copy of a message destined to it (if any has
  been received in the previous step; any choice is feasible). Confirm this to
  the original sender of the message.
  \item Delete all messages for which confirmations have been received and all
  currently held copies of messages.
  \item Set $k(i+1):=\min\{k(i)e^{\lfloor k(i)\rfloor/5}, \sqrt{\log n}\}$
  and $i:=i+1$.
\end{enumerate}
Intuitively, this algorithm exploits that the number of messages that still
needs to be delivered falls rapidly, thus enabling the nodes to try routing
increasingly many redundant copies of the remaining messages without causing
too much traffic. If just one of these copies can be deleted, the message will
not participate in the subsequent phase. Hence the number of messages will fall
by a factor that is exponential in the number of copies per message, permitting
to use an exponentially larger number of copies in the next phase without
overloading the communication network.

The techniques and proofs presented in~\cite{lenzen11} yield the following
bound on the running time of this simple algorithm for the special case of
$\Ms=\Mr=n$.
\begin{corollary}
Provided that $\Ms=\Mr=n$, the above algorithm solves Problem \ref{prob:idt} in
$\BO(\log^* n)$ synchronous rounds w.h.p., where $\log^* x$ denotes the inverse
tower function.\footnote{Formally: $\log^* x =1$ for $x\in (0,2]$ and $\log^*
x=1+ \log^* \log x$ for $x>2$. This function grows exceptionally slowly; $\log^*
x\leq 5$ for $x\leq 2^{65\,536}$.}\hfill{$\qed$}
\end{corollary}
For ease of presentation, we do not discuss asynchronicity (which is dealt with
by round counters) or the case that $\Ms,\Mr\neq n$ (requiring to adapt the
growth of $k$) here. In \cite{lenzen11} appropriate modifications of the given
algorithm are discussed, leading to the following more general
result.\footnote{In fact, \cite{lenzen11} presents an asymptotically optimal
solutions without the additive $\log^*$ overhead. However, for any practical
purposes, $\log^* n$ is a constant, and the ``optimal'' solution is more
complex, less robust, and for reasonable parameters slower than the given
algorithm.}
\begin{corollary}\label{coro:base_level}
An algorithm $\A(1)$ exists that solves Problem \ref{prob:idt} in an
asynchronous system within $\BO((\Ms+\Mr)/n+(\log^* n-\log^*(n/\Mr)))$
time w.h.p.
\end{corollary}
It is important to note that $\A(1)$ is not uniform, i.e., (an appropriate
estimate of) $\Ms+\Mr$ needs to be known to the nodes in order to execute the
algorithm. However, it is not difficult to guarantee this in a practical system
by monitoring the network load and updating the nodes frequently. Also, instead
of the ``one-shot'' version of the problem described, a perpetually running
solution is required that handles the network traffic generated over time. We
argue, however, that in light of the results from~\cite{lenzen11}, it is
feasible to study the simplified version of the problem in order to assess the
potential gain of our approach.

Having $\A(1)$ in place, we rely on recursion to solve the task on Level~$l>1$:
\begin{enumerate}
  \item Calling $\A(l-1)$, node $v\in \CX(s,l-1)_i$, $i\in\{1,\ldots,n^s\}$,
  sends each of its messages to a node in $\CX(s,l-1)_i$ whose edges in
  $E_{i,l}$ lead to the copy of $\CX(s,l-1)$ containing the destination of the
  message, choosing u.i.r.\ from the nodes fulfilling this criterion.
  \item Each node forwards the received messages over its edges in $E_{i,l}$
  to the copy of $\CX(s,l-1)$ they are destined for, balancing the load on these
  edges.
  \item $\A(l-1)$ is called again to forward all messages to their destinations.
\end{enumerate}
We will show now that this algorithm is asymptotically optimal with respect to
the number of hops (i.e., required rounds) up to a small term inherited from
$\A(1)$.
\begin{theorem}\label{theorem:algo_low_deg}
Algorithm $\A(1/s)$ solves Problem~\ref{prob:idt} on $\CX(s,1/s)$. Its running
time is w.h.p.\ bounded by $\BO((\Ms+\Mr)/n^s+(\log^* n^s-\log^*(n^s/\Mr))$.
\end{theorem}
\proof{We prove the statement by induction on $l\in \{1,\ldots,1/s\}$, i.e., we
show that for any $l$, $\A(l)$ solves Problem~\ref{prob:idt} on $\CX(s,l)$
within the stated number of rounds w.h.p. For $l=1$ this claim immediatelz
follows from Corollary~\ref{coro:base_level}. Observe that for $l>1$, $\A(l)$
will eventually deliver all messages to their destinations since $\A(1)$ does,
i.e., it is sufficient to show the stated bound on the running time. Moreover,
note that it does not matter how the constants in the $\BO$-term grow with $l$
since $l$ is constantly bounded.

Assume that the claim is correct for some $l\in \{1,\ldots,1/s-1\}$. We show
that whenever $\A(l+1)$ calls $\A(l)$, w.h.p.\ at most $\BO(\Ms+\Mr+\log n)$ many
messages have to be sent or received by any node. Recalling that message
destinations are w.l.o.g.\ distributed u.i.r., we have that the at most
$\Mr|V(\CX(s,l)|$ messages that have destinations in some given copy of
$V(\CX(s,l)$ are distributed u.i.r.\ among all copies of $V(\CX(s,l)$. Hence,
between any pair $V(\CX(s,l)_i$, $V(\CX(s,l)_j$ of such copies, in expectation
at most $\Mr|V(\CX(s,l))|/n^s$ messages need to be exchanged. Thus, Chernoff's
bound yields that w.h.p.\ no more than $\BO(\Mr|V(\CX(s,l))|/n^s+\log n)$
messages need to be sent from $V(\CX(s,l)_i$ to $V(\CX(s,l)_j$.\footnote{Note
that a simple application of the union bound shows that for any polynomial
number of events that occur w.h.p., it holds that \emph{all} of them occur
\emph{concurrently} w.h.p.}

Afterwards, applying Chernoff's bound again, we infer that the number of
messages a single node needs to receive in Step~1 is w.h.p.\ at most
$\BO(\Mr+\log n)$, as a fraction of $1/n^s$ of the nodes in $V(\CX(s,l)_i$ have
edges in $E_{i,l}$ that lead to $V(\CX(s,l)_j$. By induction hypothesis, each
call of $\A(l)$ in Step~1 will thus terminate within $\BO((\Ms+\Mr)/n^s+(\log^*
n^s-\log^*(n^s/\Mr)))$. rounds w.h.p. Moreover, as for each node its $n^s$ edges
in $E_{i,l}$ lead to the same copy of $\CX(s,l)$, Step~2 terminates in
$\BO((\Mr+\log n)/n^s)=\BO(R/n^s)$ rounds w.h.p. In addition, this implies that
no node will have to send more than $\BO(R)$ messages in Step~3 w.h.p. Hence, we
can apply the induction hypothesis again in order to see that Step~3 terminates
within $\BO((\Ms+\Mr)/n^s+(\log^* n^s-\log^*(n^s/\Mr)))$ rounds w.h.p. This
concludes the induction step and the proof.}

A closer examination of the involved constants reveals that they grow
exponentially in $1/s$. However, in recursive calls the algorithm uses
exclusively links on lower levels. Since the number of nodes on each level
grows rapidly, the physical distances of the nodes grow by more than factor 2
for each added level. Thus, overall routing path length is bounded by a
geometric series with constant limit times the lengths of links on the top
level. On the other hand, since we will choose $s$ not too small ($1/3$ resp.\
$1/4$), the number of routing hops is still small.

We remark that it is possible to generalize Theorem~\ref{theorem:algo_low_deg}
to non-constant values of $1/s\in \BO(\sqrt{\log n/\log \log n})$, however,
choosing $s$ too small is not desirable since the number of routing hops grows
exponentially in $1/s$.

\section{Performance Estimation}\label{sec:simulation}
In this section, we study the practical merits that are to be expected from
implementing the proposed communication strategy. We base our reasoning on
simulation results and discuss in detail what effects on bandwidth and delays of
arbitrary point-to-point communication can be deduced in comparison to a torus
network. Furthermore, we briefly address some advantages regarding the
robustness of the proposed routing scheme.

\subsection{Dense Traffic (Bandwidth Comparison)}\label{subsec:dense}
Theorem~\ref{theorem:algo_low_deg} states an asymptotic result, i.e., for
sufficiently large $n$ and any constant $s\in (0,1)$, outdegrees of $n^s+1/s-2$
suffice to guarantee good load balance and a running time that is only a
constant factor larger than the optimum. However, it is not clear how large the
number of nodes needs to be for a certain value of $s$ in order to ensure good
performance. The strong probability bounds obtained in
Section~\ref{sec:point-to-point} indicate that the approach is quite robust,
therefore good results for practical values of $n$ can be expected.

In order to estimate the bandwidth and delays a CLEX system will feature in
comparison to a torus grid of the same size and total bandwidth, we performed
simulations of the proposed point-to-point communication algorithm on
$\CX(1/4,4)$ with $n=32^4\approx 10^6$ and on $\CX(1/3,3)$ with $n=64^3\approx
2.5\cdot 10^5$ nodes.\footnote{\emph{Sequoia}, featuring a Blue Gene/Q
architecture, will comprise 1.6 million processors and is expected to go into
service in 2012~\cite{feldman09}.} Due to memory constraints, we confined
ourselves to simulating the algorithm synchronously and solving recursive calls
iteratively one after another. As pointed out earlier, both algorithm and
analysis are resilient to asynchronicity; hence, neither parallel nor
sequentially executed recursive calls interfere with each other, implying that
the obtained results should allow for a valid performance estimation of a
real-world system.

\begin{figure*}[tb]
  \begin{minipage}[t]{0.48\textwidth}
    \vspace{0pt}
    \centering
     \includegraphics[width=0.92\textwidth]{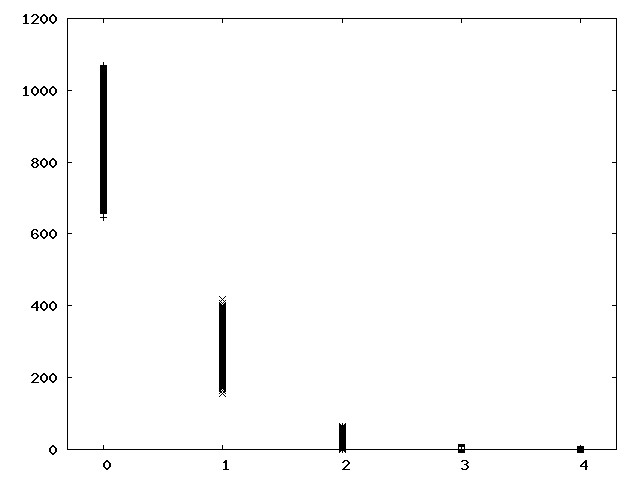}
  \end{minipage}
  \hfill
  \begin{minipage}[t]{0.48\textwidth}
    \vspace{0pt}
    \centering
    \includegraphics[width=0.92\textwidth]{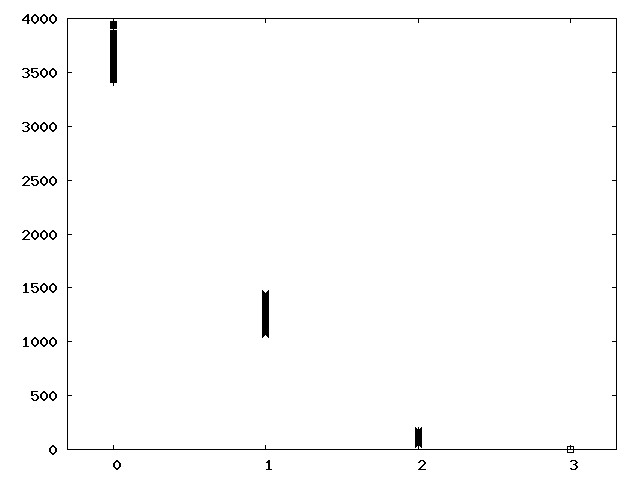}
  \end{minipage}
\caption{Number of messages instances of $\A(1)$ still needed to deliver against
passed phases. Left:~$\CX(1/4,4)$ with $32^4$ nodes and $28$ initial messages per
node, right:~$\CX(1/3,3)$ with $64^3$ nodes and $57$ initial messages per node.
In both cases most of the instances terminated after three phases; on
$\CX(1/4,4)$, a single instance required 5 phases.}
\label{fig:low_level}
\vspace*{-.4cm}
\end{figure*}

Furthermore, we adapted the algorithm from Section~\ref{sec:point-to-point}
slightly. Firstly, we are primarily interested in the case of uniformly
distributed traffic, i.e., there is no need for the algorithm to establish a
uniform distribution of messages by itself. Thus, we do not apply Valiant's
trick here, but rather start with uniformly distributed destinations. Note that
in case of ``somewhat, but not entirely uniform'' distributions, it is easy to
apply a ``lightweight'' version of Valiant's trick: just redistribute the
messages uniformly within e.g.\ level $1/s-1$ or $1/s-2$ clusters. This
drastically reduces the factor 2 overhead incurred by Valiant's trick, both with
respect to the number of hops and the distance messages travel. Secondly, for
$l>1$, in Step~2 of $\A(l)$ we choose the subset of neighbors receiving one
message more than the others uniformly at random; this slightly improves the
load balance. Thirdly, when calling $\A(1)$ on the subgraphs $\CX(1,s)$, nodes
initially send along each link one message (if available) directly to its
destination. Hence, a large fraction of the messages require only one hop to
reach their (interim) destinations. Finally, to further save bandwidth, nodes
may refrain from sending several copies of the remaining messages to potentially
relaying nodes. Rather, they merely request a message to be forwarded to its
target by a neighbor, which requires negligible\footnote{Message headers must
contain the target node ID for routing purposes (20 bits for a million nodes)
and probably some other information like e.g.\ a timestamp. Certainly the
payload of a message should be considerably larger.} $\log(32)+2 =7$
respectively $\log(64) +2= 8$ bits (the destination's identifier in the
Level-$1$ clique plus a phase counter for all phases after the first in an
asynchronous execution of algorithm $\A(1)$, cf.~Figure~\ref{fig:low_level}).
These bits may also be piggybacked on another message. Then, after receiving a
positive acknowledgement, the actual message is sent. Though this will delay the
messages that are not delivered immediately by two more rounds, we will later
see that the accordant delays do not significantly contribute to the total
time until a message is delivered.

In a first simulation experiment, we consider almost saturated channels, i.e.,
each node initially is source for $28\approx 0.9\cdot 32$ respectively $57\approx
0.9\cdot 64$ messages. Unless the communication system gets overloaded (i.e.,
more messages arrive than can be delivered quickly), it is reasonable to expect
that not all nodes generate the same amount of messages. However, due to the
randomized allocation of messages to relaying and target nodes, larger average
loads support a balanced load distribution (and thus throughput). Hence, we
chose to initially assign to all nodes roughly 90\% of the messages that can be
transferred on a level in a single round, thus permitting a worst-case estimation
of the bandwidth utilization under full load while getting useful results with
regard to message delays. Message destinations follow a uniformly random
permutation of the set containing each node $28$ resp.\ $57$ times, i.e., each
node has to send and receive the same number of messages. Thus, results afford
an easy comparison to the values for all-to-all communication given in
Section~\ref{subsec:all-to-all}.

For each Level~$l$, we measured four values: the maximal number of rounds
(excluding recursive calls) that any instance of $\A(l)$ required, the average
number of rounds (excluding recursive calls) messages spent on this level in
total, the maximal average load per node any instance of $\A(l)$ had to deal
with, and the (average) number of edges messages traversed on Level~$l$ during
the course of the complete algorithm. The outcomes of the measurements are
listed in Tables~\ref{tab:out_32_4_28} and~\ref{tab:out_64_3_57}.

\begin{table}[htb]\caption{$\A(4)$ on $\CX(1/4,4)$
with $32^4$ nodes, 28 messages from and to each node}
\begin{center}
\begin{tabular}{|c|c|c|c|c|}
\hline
\!lvl.\! & \!max.\,rds.\! & \!avg.\,rds.\! & \!max.\,avg.\,load\! &
\!avg.\,hops\!\\
\hline
1 & 11 & 13.69 & 33.44 & 10.63\\
2 & 2 & 4.11 & 30.33 & 4\\
3 & 2 & 2.05 & 28.06 & 2\\
4 & 2 & 1.03 & 28 & 1\\
\hline
\end{tabular}\label{tab:out_32_4_28}
\end{center}
\vspace*{-.4cm}
\end{table}

\begin{table}[htb]\caption{$\A(3)$ on $\CX(1/3,3)$
with $64^3$ nodes, 57 messages from and to each node}
\begin{center}
\begin{tabular}{|c|c|c|c|c|}
\hline
\!lvl.\! & \!max.\,rds.\! & \!avg.\,rds.\! & \!max.\,avg.\,load\! &
\!avg.\,hops\!\\
\hline
1 & 9 & 6.90 & 62.06 & 5.34\\
2 & 2 & 2.03 & 57.30 & 2\\
3 & 2 & 1.01 & 57 & 1\\
\hline
\end{tabular}\label{tab:out_64_3_57}
\end{center}
\vspace*{-.5cm}
\end{table}

We see that loads are well-balanced on all levels; due to the small number of
nodes on Level~$1$ some instances of $\A(1)$ invoked on $\CX(1/4,4)$ are
slightly overloaded. Accordingly, the vast majority of the messages can be
forwarded immediately on all but the first level, where a different routing
scheme is employed. On the first level, a small but relevant fraction of the
messages cannot be forwarded at once, leading to delays that are roughly 75\%
larger than the minimal possible $8$ respectively $4$ rounds.\footnote{Each
round incurs $(i)$ one ``hop'' delay $c_h$ since processors need to decide how
to deal with a message and $(ii)$ one ``propagation'' delay depending on the
respective length of connections on that level.} These messages lead to an
increase of about 30\% in traffic, since they are relayed by other nodes,
requiring one additional hop. The large maximal number of rounds $\A(1)$ takes
to complete is in accordance with theory. The algorithm runs $\BO(\log^*n)$
phases w.h.p., where in our implementation the first phase takes 1 round and
each subsequent phase 2 rounds; we incur a delay of 2 more rounds due to the
modification that relaying messages is preceded by an acknowledged request
(except for the first phase). Consequently, the algorithm should terminate
within roughly $1+2\log^* 32 = 1+2\log^* 64= 7$ rounds, which is true for most
instances; the large number of recursive calls and the fact that on $\CX(1/3,3)$
some of the calls have higher average loads than nodes' degrees explain the
differences. Figure~\ref{fig:low_level} depicts the number of remaining messages
of all invoked instances of $\A(1)$ plotted against the number of passed phases.

Note that a single message is unlikely to experience large delays in all calls 
of $\A(1)$ it participates in. The total number of rounds a message spends on 
Level~$1$ can be stochastically bounded from above by the sum of independent 
random variables describing the number of rounds passing until a message is 
forwarded in the most loaded instance of $\A(1)$. Thus, Chernoff type 
bounds 
apply, giving exponential tail bounds on the probability that the random
variable exceeds its expectation. As on higher levels almost all messages are
forwarded immediately, we have a strong indication that few messages will be
delayed more than 2-2.5 times the expected average delay, both with respect to
hops and propagation time.

Due to the increase of the number of nodes on each level by a factor of 32 (64),
the physical distances of the processors---and hence the length of connecting
cables---grow by factor $32^{1/3}\approx 3.2$ ($64^{1/3}=4$) per
level.\footnote{We assume that physical distances on Level~1 are not determined
by the volume required by the links connecting processors, but rather by cooling
requirements, i.e., the number of processors in a cube of edge length $l$ is
approximately $(l/d_{\min})^3$, where $d_{\min}$ is the minimal feasible
distance between processors. Otherwise, we had an increase of up to
$\sqrt{32/6}\approx 2.3$ ($3.3$) in cable length.} On the top level, link
lengths will be in the order of the network diameter. Shortest-path routing in a
torus grid bridges on average similar distances as one hop on the top level of
$\CX(s,1/s)$.\footnote{ Strictly speaking, the distortion of the network
embedding and the fact that messages do not take physically shortest paths
implies that on the torus topology total delays are a constant factor larger
than the average delay of top level links in $\CX(s,1/s)$. However, since we do
not quantify these influences, we do not incorporate them into our analysis.} We
conclude a \emph{worst-case} bound on the competitive ratio with respect to
$\bar{p}$ of about (cf.~Tables~\ref{tab:out_32_4_28} and~\ref{tab:out_64_3_57})
\begin{eqnarray*}
2.5 &\approx&  1.03+\frac{2.05}{3.2}+\frac{4.11}{3.2^2}+\frac{13.69}{3.2^3}\\
\mbox{resp.}\quad 2 &\approx& 1.01+\frac{2.03}{4}+\frac{6.90}{4^2}
\end{eqnarray*}
in comparison to the \emph{theoretical} optimum in a torus grid \emph{that does
not suffer from congestion}. In contrast, the average number of hop delays
decreases by factors
\begin{eqnarray*}
7.3 &\approx&  \frac{3\cdot 32^{4/3}}{2(1.03+2.05+4.11+13.69)}\\
\mbox{resp.}\quad 9.7 &\approx& \frac{3\cdot 64^3}{2(1.01+2.03+6.90)}.
\end{eqnarray*}
Recalling that delays in torus networks are dominated by the time it takes to
forward messages, we deduce that CLEX architectures will feature significantly
smaller overall message delays even when close to maximal communication
load.\footnote{We remark that our estimates do not cover a possibly increased
hop delay in the CLEX system imposed by the larger node degrees. As most hops
are inside level one clusters where $5$- or $6$-bit addresses need to be
resolved (in comparison to the three bits for grid links), one still can expect
delays that are considerably smaller.}

Next, we compare the bandwidth we provide to each node to the theoretical
optimum in a three-dimensional torus interconnection network. In both settings,
the topology appears identical to each node. Therefore, it is reasonable to
assume that each node has the same total bandwidth capacity of $B$. Moreover,
the amount of inbound and outbound communication is identical, implying that we
can confine our considerations to outgoing messages.

For symmetry reasons, in a torus network of $n=k^3$ nodes ($k$ nodes in each
spatial direction) an optimal scheme assigns each link the same capacity of
$B/6$. We partition the nodes into two sets of equal size, such that the
corresponding cut is minimum, containing $2k^2$ edges.\footnote{If the length of
cycles is different in $x$-, $y$-, and $z$-direction, we need to consider
different minimum cuts given by planes orthogonal to each dimension. It is easy
to see that for one of the cuts, the bandwidth-to-nodes ratio is at least as bad
as for the symmetric case.} If message destinations are distributed u.i.r., in
expectation every second message needs to pass this cut. Hence, with regard to
uniformly distributed traffic, the effective average bandwidth with regard to
Problem~\ref{prob:idt} provided to each node is bounded from above by
$2B/(3n^{1/3})$.

On $\CX(s,1/s)$, we assert bandwidth according to the simulation results, i.e., 
each node first divides its bandwidth to the levels according to the weights 
given by the average hops messages travelled on each 
level~(cf.~Tables~\ref{tab:out_32_4_28} and~\ref{tab:out_64_3_57}) and then the 
bandwidth on each level evenly among the links on that level. Each message will 
consume one unit of bandwidth per hop. We conclude that the gain in effective 
point-to-point bandwidth compared to the \emph{theoretical} maximum for a torus 
architecture will be at least roughly
\begin{eqnarray*}
8.6 &\approx& \frac{3\cdot 32^{4/3}}{2(1+2+4+10.63)}\\
\mbox{resp.}\quad 11.5 &\approx&  \frac{3\cdot 64}{2(1+2+5.34)}.
\end{eqnarray*}

Recall that the proposed asymmetric assignment of bandwidth also improves the 
efficiency of the simpler mechanism for all-to-all communication presented in 
Section~\ref{subsec:all-to-all}. Since most communication takes place on level 
$1$, to which we assigned the majority of the bandwidth, we achieve a bandwidth 
utilization for Problem~\ref{prob:ata} that is at least 2-competitive,
regardless of $s$.

\subsection{Light Traffic (Delay Comparison)}
Total message delays will be smaller if traffic is less dense, since most
messages can be forwarded immediately. Consequently, for a fair comparison of
delays, we consider light traffic matching the maximum throughput of a torus
network. From the previous results we infer that initial loads need to be
$4>28/8.6$ and $5>57/11.5$, respectively. Moreover, since saving bandwidth is
not crucial any longer, we can refrain from requesting message indirection on
the lowest level prior to sending complete messages. Apart from these two
modifications, the test settings are identical. The results are given in
Tables~\ref{tab:out_32_4_4} and~\ref{tab:out_64_3_5}.

\begin{table}[htb]\caption{$\A(4)$ on $\CX(1/4,4)$
with $32^4$ nodes, 4 messages from and to each node}
\begin{center}
\begin{tabular}{|c|c|c|c|c|}
\hline
\!lvl.\! & \!max.\,rds.\! & \!avg.\,rds.\! & \!max.\,avg.\,load\! &
\!avg.\,hops\!\\
\hline
1 & 5 & 9.02 & 9.02 & 10.53\\
2 & 1 & 4 & 7.32 & 4\\
3 & 1 & 2 & 4.02 & 2\\
4 & 1 & 1 & 4 & 1\\
\hline
\end{tabular}\label{tab:out_32_4_4}
\end{center}
\vspace*{-.45cm}
\end{table}

\begin{table}[htb]\caption{$\A(3)$ on $\CX(1/3,3)$
with $64^3$ nodes, 5 messages from and to each node}
\begin{center}
\begin{tabular}{|c|c|c|c|c|}
\hline
\!lvl.\! & \!max.\,rds.\! & \!avg.\,rds.\! & \!max.\,avg.\,load\! &
\!avg.\,hops\!\\
\hline
1 & 5 & 4.32 & 10.36 & 5.11\\
2 & 1 & 2 & 5.09 & 2\\
3 & 1 & 1 & 5 & 1\\
\hline
\end{tabular}\label{tab:out_64_3_5}
\end{center}
\vspace*{-.5cm}
\end{table}

We observe that dropping the mechanism to save bandwidth reduces delays on
Level~1 significantly, while due to the smaller loads the average bandwidth
consumption per message is roughly the same as before. Repeating the previous
calculations for the new data, we see that average propagation delays slightly
improve to at worst (cf.~Tables~\ref{tab:out_32_4_4} and~\ref{tab:out_64_3_5})
\begin{eqnarray*}
2.3&\approx &  1+\frac{2}{3.2}+\frac{4}{3.2^2}+\frac{9.02}{3.2^3}\\
\mbox{resp.}\quad 1.8 &\approx & 1+\frac{2}{4}+\frac{4.32}{4^2}
\end{eqnarray*}
times the average time a signal requires to follow physically shortest paths.
The required number of hops reduces considerably, widening the gap to torus
interconnection networks to factors
\begin{eqnarray*}
9.5 &\approx& \frac{3\cdot 32^{4/3}}{2(1+2+4+9.02)}\\
\mbox{resp.}\quad 13.1 &\approx & \frac{3\cdot 64^3}{2(1+2+4.32)}.
\end{eqnarray*}

Moreover, we see that all messages can be forwarded immediately on all but the
lowest level and $\A(1)$ terminates after at most 5 rounds in all instances.



\section{Conclusion}
In this work, we proposed the CLEX interconnection and routing scheme for
supercomputers. Our results emphasize the advantages of small diameters when
aiming for small delays and high bandwidth utilization in face of growing
numbers of processors. We simulated configurations of $3$- respectively
$4$-level CLEX architectures comprising half a million and a million nodes. The
results indicate performance gains of roughly an order of magnitude for
point-to-point communication in comparison to three-dimensional torus
topologies. This comparison is based on the principal limitations of a
torus topology, i.e., it does for instance not respect that a real-world
routing mechanism will not be able to concurrently propagate all messages along
shortest paths. Certainly, this performance gap will more than compensate for an
increased local switching time due to larger node degrees, and it might justify
the larger expense for the routing hardware. We believe this to be particularly
true in the future, since in the past (parallel) computation power grew much
faster than communication capacity, and there is no sign that this trend might
stop anytime soon.


\bibliographystyle{IEEEtran}
\bibliography{../sc}

%

\end{document}